\documentclass[12pt]{article}

\usepackage{graphicx}
\usepackage{amsmath}
\usepackage{amssymb}

\begin{document}

\title{Lyapunov exponents in resonance multiplets}

\author{I.~I.~Shevchenko \\
Pulkovo Observatory of the Russian Academy of Sciences \\
Pulkovskoje ave.~65, St.Petersburg 196140, Russia}
\date{}

\maketitle

\begin{abstract}
The problem of estimating the maximum Lyapunov exponents of the
motion in a multiplet of interacting nonlinear resonances is
considered for the case when the resonances have comparable
strength. The corresponding theoretical approaches are considered
for the multiplets of two, three, and infinitely many resonances
(i.e., doublets, triplets, and ``infinitets''). The analysis is
based on the theory of separatrix and standard maps. A ``multiplet
separatrix map'' is introduced, valid for description of the
motion in the resonance multiplet under certain conditions. In
numerical experiments it is shown that, at any given value of the
adiabaticity parameter (which controls the degree of
interaction/overlap of resonances in the multiplet), the value of
the maximum Lyapunov exponent in the multiplet of equally-spaced
equally-sized resonances is minimal in the doublet case and
maximal in the infinitet case. This is consistent with the
developed theory.
\end{abstract}

\medskip


\medskip


\bigskip



\bigskip

\section{Introduction}

Calculating or estimating the Lyapunov exponents provides a
powerful tool for exploring most fundamental properties of
dynamical systems in various physical and mechanical applications.
The main advantage of this tool is that it allows one to separate
chaos from order. If close trajectories in the bounded phase space
diverge exponentially, then the motion is chaotic
\cite{C79,LL92,Meiss92}. The maximum rate of this exponential
divergence is characterized by the maximum Lyapunov exponent $L$.
The quantity $T_\mathrm{L} \equiv L^{-1}$ is the so-called
Lyapunov time, representing the characteristic time of predictable
dynamics. Knowledge of the Lyapunov time allows one to judge on
the possibility for predicting the motion in chaotic domains of
phase space. Due to the exponential divergence of chaotic orbits,
the trajectory of any dynamical system cannot be accurately
predicted on timescales much greater than system's Lyapunov time;
this determines the importance of methods for estimating the
Lyapunov exponents and times in physical and mechanical
applications \cite{C79,LL92,A06}.

In this article, we consider the problem of estimating the maximum
Lyapunov exponent of the motion in a multiplet of interacting
resonances for the case when the resonances have comparable
strength. For describing nonlinear resonances, we use the
perturbed pendulum model (it was introduced in \cite{C79} as a
``universal'' one). Considering the case of interacting resonances
of comparable strength is inspired by the fact that when one
applies the perturbed pendulum model of nonlinear resonance in
various applications, one usually finds out that the perturbations
are not at all weak; see examples in \cite{S07IAU}.

\section{Resonance multiplets}
\label{mpr}

For the model of perturbed nonlinear resonance, we take the
following paradigmatic Hamiltonian \cite{S99,S00JETP}:

\vspace{-3mm}

\begin{equation}
H = {{{\cal G} p^2} \over 2} - {\cal F} \cos \phi +
    a \cos(\phi - \tau) + b \cos(\phi + \tau).
\label{h}
\end{equation}

\noindent The first two terms in Eq.~(\ref{h}) represent the
Hamiltonian $H_0$ of the unperturbed pendulum, where $\phi$ is the
pendulum angle (the resonance phase angle), and $p$ is the
momentum. The periodic perturbations are given by the last two
terms; $\tau$ is the phase angle of perturbation: $\tau = \Omega t
+ \tau_0$, where $\Omega$ is the perturbation frequency, and
$\tau_0$ is the initial phase of the perturbation. The quantities
${\cal F}$, ${\cal G}$, $a$, $b$ are constants. The frequency of
the pendulum small-amplitude oscillations is given by

\vspace{-3mm}

\begin{equation}
\omega_0 = ({\cal F G})^{1/2} .
\label{ome0}
\end{equation}

\noindent An important ``adiabaticity parameter'' \cite{C79},
measuring the relative frequency of perturbation, is

\vspace{-3mm}

\begin{equation}
\lambda = {\Omega \over \omega_0} .
\label{lambda}
\end{equation}

\begin{figure}
\begin{center}
\includegraphics[width=0.7\textwidth]{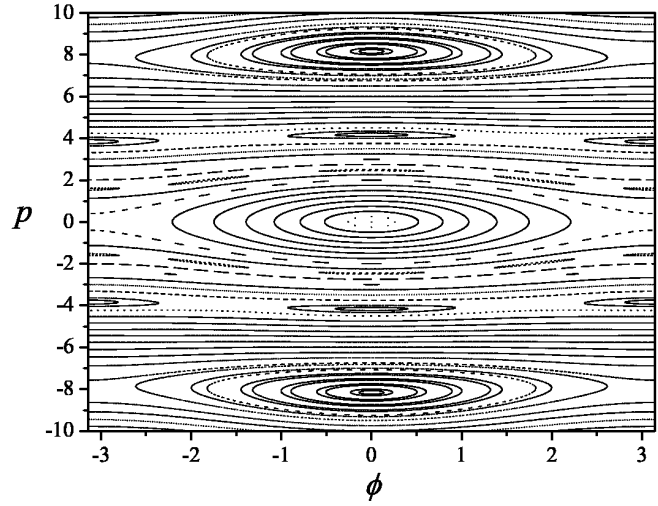}
\end{center}
\caption{\small A chaotic resonance triplet. Weak interaction
($\lambda = 8$).}
\label{fig_p8}
\end{figure}

In the well-known phase portrait ``$\phi$--$p$'' of the
non-perturbed pendulum, a single domain (``cell'') of librations,
bounded by the non-perturbed separatrix, is present. If the
perturbations are ``switched on'' (i.e., $\varepsilon \neq 0$), a
section of the phase space of motion can be constructed. Let us
construct it at $\tau = 0 \mbox{ mod } 2 \pi$, taking the
parameters' values as follows: $\Omega = 8$, $\omega_0 = 1$,
$a=b$, $\varepsilon \equiv {a \over {\cal F}} = 0.5$. The
resulting section is shown in Fig.~\ref{fig_p8}; now not one but
three domains of librations, i.e., three resonances, are present.

\begin{figure}
\begin{center}
\includegraphics[width=0.7\textwidth]{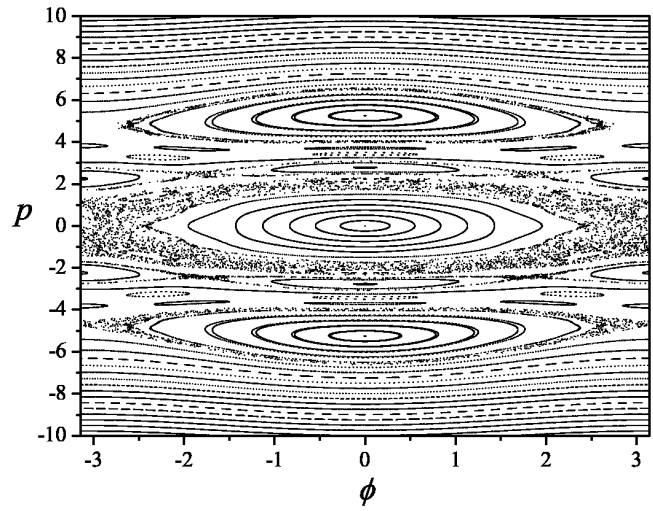}
\end{center}
\caption{\small A chaotic resonance triplet. Moderate interaction
($\lambda = 5$).}
\label{fig_p5}
\end{figure}

\begin{figure}
\begin{center}
\includegraphics[width=0.7\textwidth]{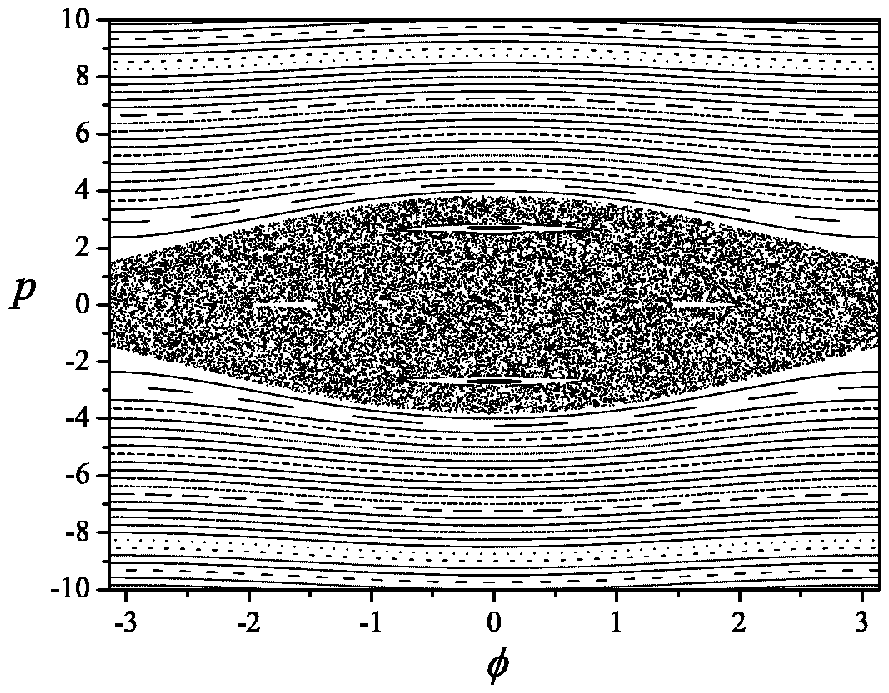}
\end{center}
\caption{\small A chaotic resonance triplet. Strong overlap
($\lambda = 2$).}
\label{fig_p2}
\end{figure}

If the perturbation frequency is relatively large (as in
Fig.~\ref{fig_p8}, where $\lambda = 8$), the separation of
resonances in the momentum $p$ is large and they almost do not
interact. On reducing the frequency of perturbation, the
resonances approach each other and appreciable chaotic layers
emerge in the vicinity of the separatrices (see Fig.~\ref{fig_p5},
where $\lambda = 5$; the value of $\varepsilon$ is as in the
previous section). As it is well visible in Fig.~\ref{fig_p5}, the
motion in the vicinity of the separatrices is irregular. On
reducing further the frequency of perturbation, the layers merge
into a single chaotic layer, due to strong overlap of the
resonances (see Fig.~\ref{fig_p2}, where $\lambda = 2$).

\section{The separatrix map}
\label{sxmap}

The chaotic layer theory has applications in various areas of
physics, mechanics and, in particular, in celestial mechanics
\cite{C79,S07IAU}. The key role in this theory is played by the
separatrix maps. They represent the motion of a system close to
separatrices in a discrete way (``stroboscopically''): system's
state, set by the ``time'' and ``energy'' variables, is mapped
discretely at the moments of passage of the positions of
equilibrium by the pendulum describing the resonance.

The motion near the separatrices of the perturbed
pendulum~(\ref{h}) with asymmetric perturbation ($a \neq b$) is
described by the so-called separatrix algorithmic map \cite{S99}:

\vspace{-3mm}

\begin{eqnarray}
& & \mbox{if } w_i < 0 \mbox{ and } W = W^-
              \mbox{ then } W := W^+, \nonumber \\
& & \mbox{if } w_i < 0 \mbox{ and } W = W^+
              \mbox{ then } W := W^-; \nonumber \\
& & w_{i+1} = w_i - W \sin \tau_i,   \nonumber \\
& & \tau_{i+1} = \tau_i + \lambda \ln {32 \over \vert w_{i+1}
\vert}
                   \ \ \ (\mbox{mod } 2 \pi);
\label{sam}
\end{eqnarray}

\noindent where $\lambda$ is given by Eq.~(\ref{lambda}), and

\vspace{-3mm}

\begin{eqnarray}
 & & W^+ (\lambda, \eta) = \varepsilon
 \lambda \left( A_{2}(\lambda) + \eta A_{2}(-\lambda) \right), \nonumber \\
 & & W^- (\lambda, \eta) = \varepsilon
 \lambda \left( \eta A_{2}(\lambda) + A_{2}(-\lambda) \right),
\label{samWpm}
\end{eqnarray}

\noindent $\varepsilon = {a \over {\cal F}}$, $\eta = {b \over
a}$. The Melnikov--Arnold integral (``MA-integral'')
$A_2(\lambda)$ is given by the formula

\vspace{-3mm}

\begin{equation}
A_2(\lambda) = 4 \pi \lambda {\exp({{\pi \lambda} / 2}) \over
\sinh (\pi \lambda)},
\end{equation}

\noindent see \cite{C79,S98PS,S00JETP}.

The quantity $w$ denotes the relative (with respect to the
separatrix value) pendulum energy: $w \equiv {H_0 \over {\cal F}}
- 1$. The variable $\tau$ is the phase angle of perturbation. One
iteration of map~(\ref{sam}) corresponds to one half-period of
pendulum's libration or one period of its rotation.

If $a = b$ (the symmetric case), the separatrix algorithmic map
reduces to the well-known ordinary separatrix map

\vspace{-3mm}

\begin{eqnarray}
& & w_{i+1} = w_i - W \sin \tau_i,  \nonumber \\
& & \tau_{i+1} = \tau_i +
                 \lambda \ln {32 \over \vert w_{i+1} \vert}
                 \ \ \ (\mbox{mod } 2 \pi),
\label{sm}
\end{eqnarray}

\noindent first written in this form in \cite{C77,C79}; the
expression for $W$ \cite{S98PS,S00JETP} is

\vspace{-3mm}

\begin{equation}
W = \varepsilon \lambda \left( A_2(\lambda) + A_2(-\lambda) \right) =
4 \pi \varepsilon {\lambda^2 \over \sinh {\pi \lambda \over 2}}.
\label{W}
\end{equation}

\noindent Formula~(\ref{W}) differs from that given in
\cite{C79,LL92} by the term $A_2(-\lambda)$, which is small for
$\lambda \gg 1$. However, its contribution is significant when
$\lambda$ is small \cite{S98PS}, i.e., in the case of adiabatic
chaos.

An equivalent form of Eqs.~(\ref{sm}), used, e.g., in
\cite{CS84,S98PLA}, is

\vspace{-3mm}

\begin{eqnarray}
     y_{i+1} &=& y_i + \sin x_i, \nonumber \\
     x_{i+1} &=& x_i - \lambda \ln \vert y_{i+1} \vert + c
                   \ \ \ (\mbox{mod } 2 \pi),
\label{sm1}
\end{eqnarray}

\noindent where $y = w / W$, $x = \tau + \pi$; and

\vspace{-3mm}

\begin{equation}
c = \lambda \ln {32 \over \vert W \vert}.
\label{c}
\end{equation}

In \cite{S00JETP}, the theory of separatrix maps was shown to be
legitimate for using to describe the motion near separatrices of
perturbed nonlinear resonances in the full range of $\lambda$,
including its low values. The half-width $y_b$ of the main chaotic
layer of the separatrix map~(\ref{sm1}) in the case of the least
perturbed border of the layer was computed as a function of
$\lambda$ in \cite[fig.~1]{S08PLA}. The observed dependence
follows a piecewise linear law with a transition point at $\lambda
\approx 1/2$. This transition takes place not only in what
concerns the width of the layer, but also in other characteristics
of the motion, in particular, in the maximum Lyapunov exponent.
The clear-cut sharp transition at this point manifests a
qualitative distinction between two types of dynamics,
``adiabatic'' (``slow'') and ``non-adiabatic'' (``fast'') chaos
\cite{S08PLA}.

The parameter $\lambda = \Omega / \omega_0$ measures the distance
between the perturbing and guiding resonances in the units of one
quarter of the width of the guiding resonance. Therefore,
$\lambda$ can be regarded as a kind of the resonance overlap
parameter \cite{S11ApJ}. It is important to note that the border
$\lambda \approx 1/2$ between the cases of adiabatic chaos and
non-adiabatic chaos does not separate the cases of resonance
overlap and resonance non-overlap: the border between the latter
cases lies much higher in $\lambda$; e.g., in the phase space of
the standard map the integer resonances start to overlap, on
decreasing $\lambda$, at $K_\mathrm{G} = 0.9716\ldots$
\cite{C79,Meiss92}, i.e., already at $\lambda = 2 \pi /
\sqrt{K_\mathrm{G}} \approx 6.37$.

\section{A multiplet separatrix map}
\label{msxmap}

Let us consider a nonlinear resonance in the perturbed pendulum
model with several harmonic perturbations (i.e., in comparison
with Hamiltonian~(\ref{h}), the number of equally-spaced
perturbing harmonics may be arbitrary):

\vspace{-3mm}

\begin{equation}
H = {{{\cal G} p^2} \over 2} - {\cal F} \cos \phi +
    \sum_{k=1}^{M} a_k \cos(\phi - k \tau) +
    \sum_{k=1}^{M} b_k \cos(\phi + k \tau).
\label{hmulti}
\end{equation}

\noindent Thus the number of resonances in the multiplet is equal
to $2M + 1$.

Let us  build a separatrix map for Hamiltonian~(\ref{hmulti}) with
the symmetric perturbations ($a_k = b_k$). If the perturbations
are asymmetric, the problem is more complicated, because the
separatrix map becomes algorithmic, as in the triplet case
\cite{S99}.

Setting $a_k = b_k$ and calculating the increment of the energy
variable (analogously to the triplet case, considered in
\cite{C79}) gives the result $\sum_{k=1}^{M} W_k \sin (k \tau_i)$,
whereas the increment of the time variable remains the same as in
the triplet case. Thus the separatrix map~(\ref{sm}) is
generalized to a ``multiplet separatrix map'', given by

\vspace{-3mm}

\begin{eqnarray}
& & w_{i+1} = w_i - \sum_{k=1}^{M} W_k \sin (k \tau_i) ,
\nonumber \\
& & \tau_{i+1} = \tau_i + \lambda \ln {32 \over \vert w_{i+1} \vert}
                 \ \ \ (\mbox{mod } 2 \pi),
\label{msm}
\end{eqnarray}

\noindent where

$$
W_k = 4 \pi \varepsilon_k \frac{\lambda_k^2}{\sinh \frac{\pi
\lambda_k}{2}} ,
$$

\noindent where $\lambda_k = k \lambda$ and $\varepsilon_k \equiv
{a_k \over {\cal F}} = {b_k \over {\cal F}}$.

The domain of validity of map~(\ref{msm}) (in describing the
near-separatrix motion) is expected to be usually much smaller
than that of map~(\ref{sm}), because the natural condition of
validity $\vert W \vert \lesssim 1$ generalizes here to the
condition $\sum_{k=1}^{M} \vert W_k \vert \lesssim 1$. Thus, if
there is a lot of perturbing harmonics, the maximum allowed
amplitudes $\varepsilon_k$ in the multiplet case must be usually
much smaller than the maximum allowed amplitude $\varepsilon$ in
the triplet case, at any given value of $\lambda$.

Also note that in the case of non-adiabatic perturbation ($\lambda
\gtrsim 1/2$) the multiplet map~(\ref{msm}) can be usually
replaced by the classical map~(\ref{sm}) for the ``central''
triplet (with $W = W_1$), because at high values of $\lambda$ the
coefficients $W_k$ at $k > 1$ are exponentially small with $k$,
with respect to $W_1$.

\section{Analytical estimating the Lyapunov exponents}
\label{iau_maelt}

The maximum Lyapunov exponent is defined by the formula

\begin {equation}
L = \limsup _ {{t \to \infty} \atop {d (t_0) \to 0}} {1 \over
{t-t_0}} \ln {d (t) \over d (t_0)} ,
\label {def_lce}
\end {equation}

\noindent where $d(t_0)$ is the distance (in the phase space of
motion) between two nearby initial conditions for two trajectories
at the initial instant of time $t_0$, and $d(t)$ is the distance
between the evolved initial conditions at time $t$ (e.g.,
\cite{LL92}).

The art of calculation of the Lyapunov exponents (and, in
particular, the maximum Lyapunov exponent) on computers has more
than a thirty-year history and during this time it has become an
extensive part of applied mathematics; see reviews in
\cite{F84,LL92}. Modern numerical methods for computation of the
Lyapunov exponents are effective and precise. Approaches for
analytical estimating the Lyapunov exponents were started to be
developed relatively recently, beginning with those providing
precision by the order of magnitude \cite{MF96,NM99}, and later on
providing precision comparable to the numerical methods
\cite{HM96,MH97,S00GAO,S02CR,S07IAU,S08MN}, though in limited
applications.

Morbidelli and Froeschl\'e \cite{MF96} and Nesvorn\'y and
Morbidelli \cite[p.~256]{NM99} suggested to estimate the Lyapunov
time by taking it equal, by the order of magnitude, to the
libra\-tion/cir\-culation period of the resonant angle, or, in
practice, to the period of small-amplitude oscillations on
resonance (i.e., $\sim 1/\omega_0$). This estimate has a rather
limited domain of validity; in fact, as we shall see in
Section~\ref{sec_tvne}, $L \sim \omega_0$ solely at $\lambda
\backsimeq 1$, and the ratio $L/\omega_0 \to 0$ in both limits
$\lambda \to 0$ and $\lambda \to \infty$; besides, $L/\omega_0$
rather strongly depends on other parameters, such as the
perturbation amplitude $\varepsilon$.

A different approach, based on derivation of a discrete map for a
triplet, was proposed by Holman and Murray \cite{HM96,MH97}. This
is also a one-parameter approach, but using an effective resonance
overlap parameter $K_\mathrm{eff}$ instead of $\omega_0$ or
$\Omega$. The strength of perturbation is ignored, only
frequencies are taken into account. In some way $K_\mathrm{eff}$
is analogous to the stochasticity parameter $K$ of the standard
map (whose theory is given in \cite{C79}),
though $K_\mathrm{eff}$ was introduced for a triplet. Holman and
Murray derived heuristic formulas for estimating the maximum
Lyapunov exponent in the case of moderate resonance overlap,
when $K_\mathrm{eff} \sim 1$, and in the case of strong overlap
(the adiabatic case), when $K_\mathrm{eff} \gg 1$. In the first
case, the maximum Lyapunov exponent was estimated in \cite{HM96}
as $L \approx \omega_0$ (the frequency of small oscillations on
the resonance), and in the second case as $L \approx \Omega$ (the
frequency of external perturbation). Murray and Holman \cite{MH97}
refined somewhat the formula in the case of strong overlap
($K_\mathrm{eff} \gg 1$) by introducing a logarithmic dependence
on $K_\mathrm{eff}$, namely $L \propto \ln (K_\mathrm{eff} / 2)$,
the function essentially the same as for the standard map (see
Subsection~\ref{lemsm}), though derived for a triplet. For the
whole range of resonance overlap, $1 < K_\mathrm{eff} < +\infty$,
they proposed the following interpolating formula\footnote{There
is a misprint in the original paper \protect\cite{MH97}. We quote
the corrected formula, as given in
\protect\cite[Eq.~(12.14)]{Morbi02}.}:

\begin {equation}
L = {\Omega \over 2 \pi} \ln \left( 1 + {K_\mathrm{eff} \over 4} +
           \left( {K_\mathrm{eff} \over 2} +
                 \left( {K_\mathrm{eff} \over 4} \right)^2
           \right)^{1/2}
    \right) .
\label{MH97}
\end {equation}

\noindent Let $K_\mathrm{eff} = 1$ (this value belongs to the case
of moderate overlap), then, taking $\Omega = 2 \pi$, one has
$L=\ln 2 \approx 0.69$. For the standard map, the actual value of
$L$ at $K=1$ and $\Omega = 2 \pi$ is $\approx 0.13$ (as we shall
see in Subsection~\ref{lemsm}, Eq.~(\ref{Lstmla1})), and for the
triplet it is smaller. Thus Eq.~(\ref{MH97}) can be used for
estimates solely by the order of magnitude, because the
perturbation strength, asymmetry of perturbation, and number of
resonances in multiplets are ignored in it. It cannot be used in
the case of weak interaction of resonances, when they do not
overlap.

In \cite{S00GAO,S02CR}, an approach for estimating the maximum
Lyapunov exponent of the chaotic motion in the vicinity of
separatrices of a perturbed nonlinear resonance was proposed in
the framework of the separatrix map theory. We follow the approach
\cite{S00GAO,S02CR}, representing the maximum Lyapunov exponent
$L$ of the motion in the main chaotic layer of system~(\ref{h}) as
the ratio of the maximum Lyapunov exponent $L_\mathrm{sx}$ of its
separatrix map and the average period $T$ of rotation (or,
equivalently, the average half-period of libration) of the
resonance phase $\phi$ inside the layer. For convenience, we
introduce the non-dimensional quantity $T_\mathrm{sx} = \Omega T$.
Then the general expression for $L$ is

\vspace{-3mm}

\begin{equation}
L = \Omega {L_\mathrm{sx} \over T_\mathrm{sx}}. \label{lceh}
\end{equation}

\noindent The quantity $T_\mathrm{L} \equiv L^{-1}$, by
definition, is the Lyapunov time.

In \cite{S07IAU}, the following four generic kinds of interacting
resonances were considered: fast-chaotic resonance triplet,
fast-chaotic resonance doublet, slow-chaotic resonance triplet,
and slow-chaotic resonance doublet. Here we present formulas for
the Lyapunov time $T_\mathrm{L}$ \cite{S07IAU,S08MN} for these
four cases, and then proceed to considering a fifth generic kind,
that of infinitely many interacting resonances.

\subsection{Fast chaos. Resonance triplet}
\label{fcrt}

Assume that $a=b$ and $\lambda > 1/2$ in Eq.~(\ref{h}). Then one
has a symmetric triplet of interacting resonances, and chaos is
non-adiabatic. Following \cite{S00GAO, S02CR}, we take the
$\lambda$ dependence of the maximum Lyapunov exponent of the
separatrix map~(\ref{sm1}) in the form

\vspace{-3mm}

\begin{equation}
L_\mathrm{sx}(\lambda) \approx C_h {2 \lambda \over 1 + 2
\lambda}, \label{Lsx}
\end{equation}

\noindent where $C_h \approx 0.80$ is Chirikov's constant
\cite{S04JETPL}. The average increment of $\tau$ (proportional to
the average libration half-period, or rotation period) in the
chaotic layer is \cite{C79, S00GAO, S02CR}:

\vspace{-3mm}

\begin{equation}
T_\mathrm{sx}(\lambda, W) \approx \lambda \ln {32 e \over \lambda
| W |}, \label{Tsx}
\end{equation}

\noindent where $e$ is the base of natural logarithms.

Then, the Lyapunov time for the fast-chaotic resonance triplet
\cite{S07IAU} is given by

\vspace{-3mm}

\begin{equation}
T_\mathrm{L} = {T_\mathrm{pert} \over 2 \pi} {T_\mathrm{sx} \over
L_\mathrm{sx}} \approx T_\mathrm{pert} {(1 + 2 \lambda) \over 4
\pi C_h} \ln {32 e \over \lambda | W |}, \label{TLft}
\end{equation}

\noindent where $T_\mathrm{pert} = 2 \pi / \Omega$ is the period
of perturbation.

\subsection{Fast chaos. Resonance doublet}
\label{fcrd}

In the completely asymmetric case, when $a = 0$ or $b = 0$, the
maximum Lyapunov exponent can be found by averaging the
contributions of all separate components of the chaotic layer
\cite{S07IAU}. The averaged (over the whole layer) value of the
maximum Lyapunov exponent is the sum of weighted contributions of
the layer components corresponding to librations, direct rotations
and reverse rotations of the model pendulum. The weights are
directly proportional to the times that the trajectory spends in
the components, and, due to the supposed approximate ergodicity,
to the relative measures of the components in the phase space.
Then, the formula for the Lyapunov time for the fast-chaotic
resonance doublet \cite{S07IAU,S11ApJ} is given by

\vspace{-3mm}

\begin{equation}
T_\mathrm{L} \approx
\displaystyle{
\frac{T_\mathrm{pert}}{2 \pi} \cdot
\frac{\mu_\mathrm{libr} + 1} {\mu_\mathrm{libr}
\frac{L_\mathrm{sx}(2 \lambda)}{T_\mathrm{sx}(2 \lambda, W)} +
\frac{L_\mathrm{sx}(\lambda)}{T_\mathrm{sx}(\lambda, W)}}
} ,
\label{TLfd}
\end{equation}

\noindent where $\mu_\mathrm{libr} \approx 4$, and $W$,
$L_\mathrm{sx}$, and $T_\mathrm{sx}$ are given by
formulas~(\ref{W}), (\ref{Lsx}), and (\ref{Tsx}).

\subsection{Slow chaos. Resonance triplet}
\label{scrt}

If $\lambda < 1/2$, the diffusion across the chaotic layer is
slow, and on a short time interval the trajectory of the
separatrix map~(\ref{sm1}) follows close to some current
``guiding'' curve \cite{S07IAU,S08MN}, and this allows one to
estimate characteristics of the chaotic layer in a straightforward
manner; in particular, the Lyapunov time for this resonance type
is given by

\vspace{-3mm}

\begin{equation}
T_\mathrm{L} \approx  {T_\mathrm{pert} \over 2 \pi} \left( \ln
\left| 4 \sin \frac{c}{2} \right| + \frac{c}{\lambda} \right) ,
\label{TLst1}
\end{equation}

\noindent where $c = \lambda \ln \frac {32}{|W|}$ (Eq.~(\ref{c})).
This formula has specific limits of applicability \cite{S08MN},
namely, the parameter $c$ (approximately equal to $\lambda \ln
\frac{4}{\lambda | \varepsilon |}$ in the adiabatic case) should
not be close to $0 \mbox{ mod } 2 \pi$.

At $\lambda \ll 1$ one has $W \approx 8 \varepsilon \lambda$,
hence the approximate formula for the Lyapunov time is

\vspace{-3mm}

\begin{equation}
T_\mathrm{L} \approx  {T_\mathrm{pert} \over 2 \pi} \ln \left|
\frac{16}{\varepsilon \lambda} \sin \left( \frac{\lambda}{2} \ln
\frac{4}{ |\varepsilon| \lambda} \right) \right|. \label{TLst}
\end{equation}

\subsection{Slow chaos. Resonance doublet}
\label{scrd}

In this case, the separatrix algorithmic map~(\ref{sam})
degenerates to the ordinary separatrix map~(\ref{sm}) with $W
\approx 4 \varepsilon \lambda$, i.e., mathematically this case is
equivalent to the case of slow-chaotic resonance triplet, but with
a different (halved) value of $W$ \cite{S07IAU,S08MN}. The
Lyapunov time is then given by

\vspace{-3mm}

\begin{equation}
T_\mathrm{L} \approx  {T_\mathrm{pert} \over 2 \pi} \ln \left|
\frac{32}{\varepsilon \lambda} \sin \left( \frac{\lambda}{2} \ln
\frac{8}{ |\varepsilon| \lambda} \right) \right| , \label{TLsd}
\end{equation}

\noindent provided that the parameter $c$ is not close to $0
\mbox{ mod } 2 \pi$.

\subsection{Lyapunov exponents in supermultiplets.
The standard map theory}
\label{lemsm}

Let us assume that the number of resonances in a resonance
multiplet is greater than~3. In applications, this number can be
very large \cite{Morbi02}; then, the multiplet is called a
``supermultiplet''. If chaos is non-adiabatic ($\lambda \gtrsim
1/2$), then one can apply, as an approximation, the formulas given
in Subsections~\ref{fcrt} and \ref{fcrd} for the triplet and
doublet (depending on the perturbation asymmetry) cases, because
the influence of the ``far away'' resonances is exponentially
small with $\lambda$. However, if chaos is adiabatic ($\lambda
\lesssim 1/2$), the triplet or doublet approximations do not work
and one has to develop a different approach. Let us consider a
limiting case, namely, the case of infinitely many interacting
equally-sized equally-spaced resonances.

The standard map

\vspace{-6mm}

\begin{eqnarray}
y_{i+1} &=& y_i + K \sin x_i \ \ \ (\mbox{mod } 2 \pi), \nonumber \\
x_{i+1} &=& x_i + y_{i+1} \ \ \ (\mbox{mod } 2 \pi)
\label{stm2}
\end{eqnarray}

\noindent describes the motion in an infinite multiplet of
equally-sized equally-spaced resonances, as it is clear from its
Hamiltonian \cite{C79}:

\vspace{-3mm}

\begin{equation}
H = \frac{y^2}{2} + \frac{K}{(2 \pi)^2} \sum_{k=-N}^N \cos(x - k t) , \label{h_stm2}
\end{equation}

\noindent where $N = \infty$. The variables $x_i$, $y_i$ of
map~(\ref{stm2}) correspond to the variables $x(t_i)$, $y(t_i)$ of
the continuous system~(\ref{h_stm2}) taken stroboscopically at
time moduli $2 \pi$ (see, e.g., \cite{C79}).

\begin{figure}[t!]
\centering
\includegraphics[width=0.7\textwidth]{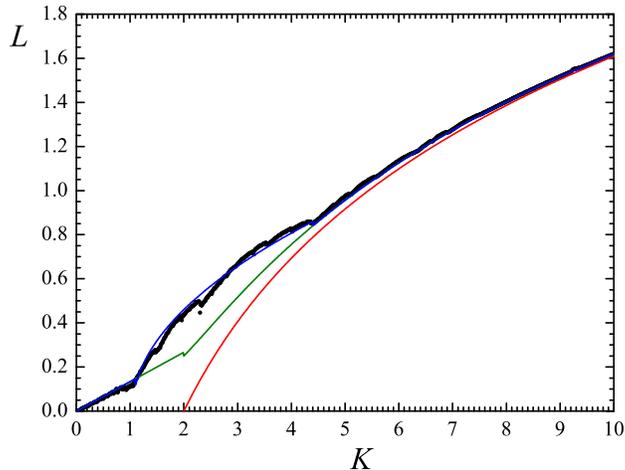} \\
\caption{The dots show the numerical-experimental dependence
$L(K)$ for the standard map at $0<K<10$, according to
\protect\cite{S04PLA,S04JETPL}. The lower curve shows the function
$\ln{\frac{K}{2}}$. The middle curve shows the glued
functions~(\protect\ref{Lstmla}), and the upper curve is given by
Eq.~(\protect\ref{Lstmla1}), where $T_\mathrm{pert}=1$.}
\label{msm_fig2}
\end{figure}

The asymptotic formula for the maximum Lyapunov exponent of the
standard map at $K \gg 1$ was derived in \cite{C79}: $L \propto
\ln \frac{K}{2}$. Rather precise fitting formulas were obtained in
\cite{S04PLA,S04JETPL} for the $L(K)$ dependence at $K < 1$ and $K
> 4.5$:

\vspace{-3mm}

\begin{equation}
L = \frac{1}{T_\mathrm{pert}} \cdot
\begin{cases}
0.1333 K, & \text{if $K < 1$}, \\
\ln \frac{K}{2} + \frac{1}{K^2}, & \text{if $K > 4.5$},
\end{cases}
\label{Lstmla}
\end{equation}

\noindent where $K = (2 \pi /\lambda)^2$. The
functions~(\ref{Lstmla}) are depicted in Fig.~\ref{msm_fig2}. In
this plot, they are glued at $K=2$; this trick apparently results
in underestimating the actual values of $L$ in the interval $1
\lesssim K \lesssim 4.5$. Arranging a better fit for $L(K)$ at
this interval, one arrives at the formulas

\vspace{-3mm}

\begin{equation}
L = \frac{1}{T_\mathrm{pert}} \cdot
\begin{cases}
0.1333 K , & \text{if $K < 1.1$}, \\
0.469 (K - 1.037)^{1/2} , & \text{if $1.1 \leq K < 4.4$}, \\
\ln \frac{K}{2} + \frac{1}{K^2} , & \text{if $K \geq 4.4$},
\end{cases}
\label{Lstmla1}
\end{equation}

\noindent which describe the behavior of $L(K)$ at $1 \lesssim K
\lesssim 4.5$ (corresponding to $3.0 \lesssim \lambda \lesssim
6.3$) much more accurately.

Thus the Lyapunov time in the ``infinitet'' case is given by

\vspace{-3mm}

\begin{equation}
T_\mathrm{L} \approx T_\mathrm{pert} \cdot
\begin{cases}
\displaystyle{\frac{7.50}{K}} (\approx 0.190 \lambda^2), &
\text{if $K < 1.1$ (or, if $\lambda > 6.0$)}, \\
2.133 (K - 1.037)^{-1/2} , &
\text{if $1.1 \leq K < 4.4$ (or, if $3.0 < \lambda \leq 6.0$)}, \\
\displaystyle{\left( \ln \frac{K}{2} + \frac{1}{K^2} \right)^{-1}} , &
\text{if $K \geq 4.4$ (or, if $\lambda \leq 3.0$)},
\end{cases}
\label{TLinf}
\end{equation}

\noindent where

\vspace{-3mm}

\begin{equation}
K = (2 \pi /\lambda)^2 .
\label{Kla}
\end{equation}

\begin{figure}[t!]
\centering
\includegraphics[width=0.7\textwidth]{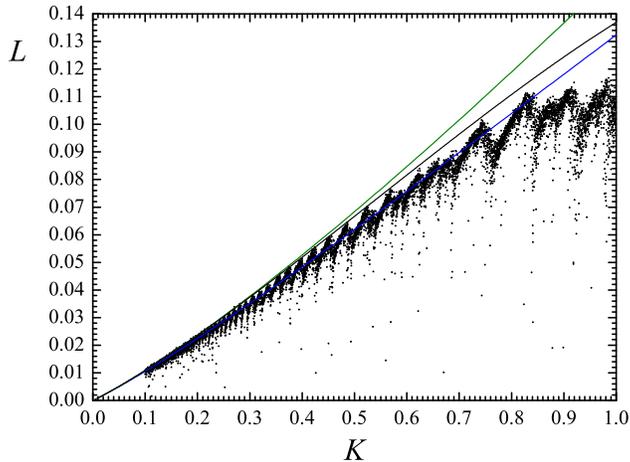} \\
\caption{The numerical-experimental $L(K)$ dependence (dots) for
the standard map \protect\cite{S04PLA,S04JETPL} at $0<K<1$. The
lower solid curve is given by the separatrix map theory without
any correction to the MA-integral; the upper solid curve is given
by the separatrix map theory with the Chirikov zero-order
correction to the MA-integral; the middle solid curve is given by
the separatrix map theory with the Chirikov--Lazutkin--Gelfreich
correction to the MA-integral.} \label{msm_fig1}
\end{figure}

A well-known important constant of the standard map dynamics is
the critical value of the parameter $K$, namely, $K_G =
0.971635406\ldots$; see, e.~g., \cite{Meiss92}. It is obvious from
Figs.~\ref{msm_fig2} and \ref{msm_fig1}, that at $K \lesssim 1$,
i.e., at $K$ below its approximate critical value, the dependence
$L(K)$, if smoothed, is close to linear. This is explainable in
the framework of the separatrix map theory \cite{S04PLA}. Indeed,
one can find the maximum Lyapunov exponent here using
formula~(\ref{TLft}) for the fast-chaotic resonance triplet,
because at $K \lesssim 1$ one has $\lambda \gtrsim 6$ and
therefore the perturbing resonances non-neighboring the guiding
one can be ignored in the first approximation (their contribution
is considered below). Thus $L = \Omega L_\mathrm{sx} /
T_\mathrm{sx}$ (Eq.~(\ref{lceh})), where $L_\mathrm{sx}$ is given
by Eq.~(\ref{Lsx}) and $T_\mathrm{sx}$ is given by
Eq.~(\ref{Tsx}). As follows from Eq.~(\ref{Lsx}), $L_\mathrm{sx}$
is practically constant at $\lambda \gtrsim 6$. On the other hand,
$T_\mathrm{sx}$ is directly proportional to $K^{-1}$ at small
enough values of $K$, as follows from Eq.~(\ref{Tsx}) (or see
Eq.~(6.18) in~\cite{C79}). Therefore, $L \propto K$ at small
enough values of $K$. However, this linear asymptotic behavior has
a slope somewhat less than the average one adopted in
approximation~(\ref{Lstmla1}), where $L \approx 0.1333 K$. Indeed,
a careful inspection of Fig.~\ref{msm_fig1} indicates that the
slope of the smoothed dependence decreases with $K$.

Let us derive a formula for $L(K)$ at $0 \leq K \lesssim 1$. This
will be a formula for the upper envelope of the observed
``ragged'' dependence (which has sharp local minima due to
marginal resonances at the borders of the chaotic layer), because
our theory (described in Section~\ref{fcrt}) is valid in the
absence of marginal resonances. (The role of marginal resonance in
defining the width of the chaotic layer is described in
\cite{S98PS,S12PRE,So12}.)

We proceed from the basic relation~(\ref{lceh}) $L = \Omega
{L_\mathrm{sx} \over T_\mathrm{sx}}$, where $\Omega = 2 \pi$,
$L_\mathrm{sx}$ is given by Eq.~(\ref{Lsx}), and $T_\mathrm{sx}$
is given by Eq.~(\ref{Tsx}). However, we modify the expression for
$W$, which enters in Eq.~(\ref{Tsx}), changing $W$ to
$W_\mathrm{st} = R_\mathrm{st} W$, where $R_\mathrm{st}$ is a
correction factor, introduced by Chirikov \cite{C79} to account
for specific properties of the standard map. Thus the formula for
$T_\mathrm{sx}$ attains the form

\begin{equation}
T_\mathrm{sx} = \lambda \ln \frac{32 e}{\lambda R_\mathrm{st}
\vert W \vert} .
\label{Tsxst}
\end{equation}

\noindent Expressing $\lambda$ through $K$, one arrives at a
formula, derived in \cite{C79} for the average half-period of
librations (or, the average period of rotations) in the chaotic
layer of the integer resonance of the standard map; this formula
is as follows:

\begin{equation}
T_\mathrm{sx} = \Omega \left( \frac{\pi^2}{K} - K^{-1/2} \ln
\frac{2 R_\mathrm{st} \pi^4}{e K^{3/2}} \right) .
\label{TsxstK}
\end{equation}

The introduction of the correction factor $R_\mathrm{st}$ is
necessary for the separatrix-map correct description of the
chaotic layer of the integer resonance of the standard map.
Chirikov's numerical-expe\-rimental estimate of the correction
factor gave $R_\mathrm{st} \approx 2.15$~\cite{C79}. Later on,
this factor was found out \cite{VC98,V99} to be expressed through
the so-called Lazutkin splitting constant: $R_\mathrm{st} = f_0
/(16 \pi^3) \approx 2.2552$, where the Lazutkin constant $f_0 =
1118.8277059409008 \dots$.

At non-zero $K$, the stable and unstable separatrices of the
integer resonance of the standard map intersect transversally;
Lazutkin \cite{L05} obtained an asymptotic (at $K \ll 1$) formula
for the separatrix splitting angle. The splitting angle at the
first intersection of the separatrices with the line $x = \pi$ is
given by

\begin{equation}
\alpha = \frac{\pi}{h^2} \exp \left( -\frac{\pi^2}{h} \right)
\sum_{m=0}^\infty c_m h^{2 m} ,
\label{alpha}
\end{equation}

\noindent where

\begin{equation}
h = \ln \left(1 + \frac{K}{2} + \left(K + \frac{K^2}{4}
\right)^{1/2} \right) ,
\label{hsxst}
\end{equation}

\noindent and the first three coefficients $c_m$ are given by the
formulas

\begin{equation}
c_0 = f_0 , \quad c_1 = f_1 - \frac{c_0}{4} , \quad c_2 = f_2 -
\frac{c_1}{4} - \frac{25 c_0}{72} ,
\label{csxst}
\end{equation}

\noindent where

\vspace{-5mm}

\begin{equation}
f_0 = 1118.8277059 \dots , \quad f_1 = 18.59891 \dots , \quad
f_2 = -2.17205 \dots .
\label{fsxst}
\end{equation}

\noindent \cite{L05,G99}. Taking into account the asymptotic
expansion~(\ref{alpha}), one arrives at

\begin{equation}
R_\mathrm{st} \approx \frac{1}{16 \pi^3}(c_0 + c_1 h^2 + c_2 h^4)
, \label{Rsxst}
\end{equation}

\noindent where $h \approx K^{1/2}$.

Combining Eqs.~(\ref{lceh}), (\ref{Lsx}), (\ref{Tsxst}), and
(\ref{Rsxst}), we build a theoretical $L(K)$ curve; it is the
middle solid one in Fig.~\ref{msm_fig1}. For comparison, the lower
solid curve in this Figure is given by the separatrix map theory
without any correction to the MA-integral (i.e., $R_\mathrm{st} =
1$), and the upper solid curve is given by the separatrix map
theory with the Chirikov zero-order (in $h$) correction to the
MA-integral (i.e., $R_\mathrm{st} = 2.2552$). One can see that the
middle curve, built on the basis of the most refined theory,
provides the best approximation for the upper envelope of the
numerical-experimental relationship, as expected.

\section{Theory versus numerical experiment}
\label{sec_tvne}

In this Section we verify our theoretical results versus numerical
simulations. For computing the maximum Lyapunov exponent (and,
generally, the Lyapunov spectra) we use the algorithms and
software developed in \cite{SK02,KS05} on the basis of the HQRB
numerical method by von~Bremen et al.\ \cite{BUP97} for
calculation of the Lyapunov spectra. The HQRB method is based on
the QR decomposition of the tangent map matrix using the
Householder transformation. For computing the trajectories we use
the integrator by Hairer et al.\ \cite{HNW87}, realizing an
explicit 8th order Runge--Kutta method (with the step size
control) due to Dormand and Prince.

\begin{figure}[t!]
\centering
\includegraphics[width=0.7\textwidth]{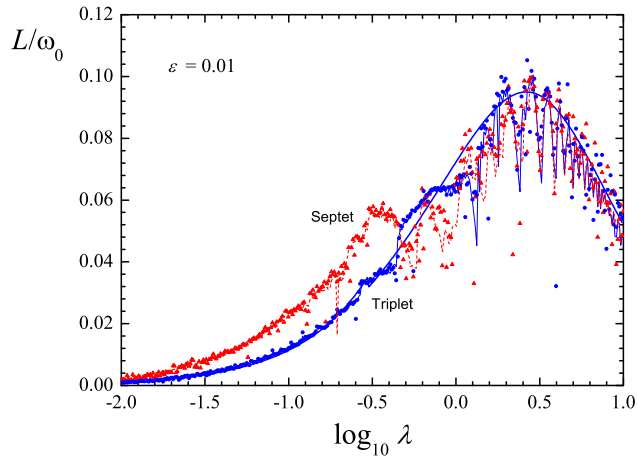} \\
\caption{The $\lambda$ dependences of the maximum Lyapunov
exponent, normalized by $\omega_0$, in the triplet and septet
cases; $\varepsilon_k = \varepsilon = 0.01$. The dots and
triangles show the numerical-experimental data obtained for the
triplet and septet, respectively, by means of numerical
integrations of the equations of motion. The thin solid and dashed
curves show the numerical-experimental data obtained by iterations
of the multiplet separatrix map for the triplet and septet,
respectively. The thick solid curve represents the separatrix map
theory (given by Eqs.~(\protect\ref{TLft}) and
(\protect\ref{TLst})) for the triplet. } \label{msm_fig3a}
\end{figure}

Let us consider first of all a small perturbation amplitude,
namely, we set $\varepsilon_k = \varepsilon = 0.01$ in
Eq.~(\ref{hmulti}). The corresponding $\lambda$ dependences of the
maximum Lyapunov exponent, normalized by $\omega_0$, are shown in
Fig.~\ref{msm_fig3a} for the triplet case ($M=1$ in
Eq.~(\ref{hmulti})) and for the septet case ($M=3$ in
Eq.~(\ref{hmulti})). The dots and triangles denote the
numerical-experimental data obtained for the triplet and septet,
respectively. The thin curves show the numerical-expe\-rimental
data obtained by iterations of the multiplet separatrix
map~(\ref{msm}), solid and dashed for the triplet and septet,
respectively. The thick solid curve represents the separatrix map
theory (given by Eqs.~(\protect\ref{TLft}) and
(\protect\ref{TLst})) for the triplet. One can see that the theory
is impressively good for the triplet. No theory is yet available
for the septet; however, the multiplet separatrix map data and the
results of direct numerical integrations are in obviously good
agreement. At $\lambda \gtrsim 0.5$, i.e., in the domain of
non-adiabatic chaos, the theory for the fast-chaotic triplet works
good for both triplet and septet, because the perturbing role of
the harmonics farther than the neighbors of the guiding resonance
is negligible.

Now let us consider the ultimately large perturbation amplitude,
namely, $\varepsilon_k = \varepsilon = 1$; in other words, let us
consider equally-sized equally-spaced multiplets. We call the
amplitude $\varepsilon = 1$ ultimately large, because the case of
$\varepsilon > 1$ can be reduced to the case of $\varepsilon < 1$
by changing the choice of the guiding resonance.

The standard map theory, given by formulas~(\ref{Lstmla1}) and
(\ref{TLinf}), can be presumably applied for estimating the
maximum Lyapunov exponents in multiplets of equally-sized
equally-spaced resonances, when the number of resonances is large,
assuming that the limiting case $M=\infty$ describes the situation
at $M \gg 1$.

\begin{figure}[t!]
\centering
\includegraphics[width=0.7\textwidth]{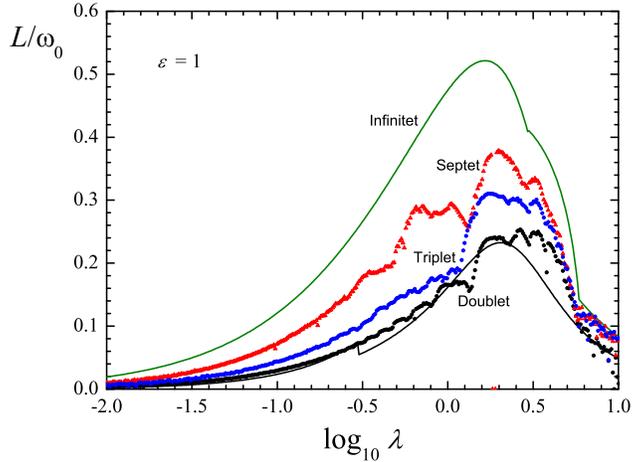} \\
\caption{The $\lambda$ dependences of the maximum Lyapunov
exponent, normalized by $\omega_0$, for multiplets of
equally-sized equally-spaced resonances. The dots show
numerical-experimental data, and the curves show theoretical
functions. The upper solid curve, given by
Eqs.~(\protect\ref{Lstmla1}) and (\protect\ref{Kla}), represents
the standard map theory for the infinitet; and the lower solid
curve, given by Eqs.~(\protect\ref{TLfd}) and
(\protect\ref{TLsd}), represents the separatrix map theory for the
doublet.} \label{msm_fig3}
\end{figure}

The $\lambda$ dependences, both theoretical and
numerical-experimental, of the maximum Lyapunov exponent
(normalized by $\omega_0$) for several multiplets of equally-sized
equally-spaced resonances are shown in Fig.~\ref{msm_fig3}. One
can see that the dependence for the septet occupies an
intermediate (in the vertical axis) position between the
dependence for the doublet and the dependence for the
``infinitet'', i.e., for the standard map. The numerical data for
the doublet agrees well with the separatrix map theory presented
in Subsections~\ref{fcrd} and \ref{scrd}, notwithstanding the
large perturbation amplitude $\varepsilon = 1$.

Comparing the heights of the curves maxima in Fig.~\ref{msm_fig3a}
(where $\varepsilon = 0.01$) and Fig.~\ref{msm_fig3} (where
$\varepsilon = 1$), one can see that $L/\omega_0$ depends strongly
on the perturbation amplitude $\varepsilon$, the difference being
obvious (about three times). This emphasizes the fact that taking
into account solely the frequencies $\Omega$ and $\omega_0$ is
insufficient for analytical estimates of $L$: the perturbation
strength must be also taken into account whenever more or less
precise estimates of $L$ are sought for.

Inspecting the plots in Fig.~\ref{msm_fig3} allows one to
qualitatively estimate the relative range of the Lyapunov exponent
values between the doublet, triplet and infinitet cases. Let us
designate the Lyapunov exponents for these three cases as
$L^{(2)}$, $L^{(3)}$, and $L^{(\infty)}$, respectively. One can
see that at $\lambda \sim 1$--3, i.e., where the values of
$L/\omega_0$ are maximal, the ratios $L^{(\infty)}/L^{(2)}$ and
$L^{(\infty)}/L^{(3)}$ are of the order of 2. At the maxima of the
curves, they are equal to 2.2 and 1.7, respectively.

It is also of interest how do the ratios $L^{(\infty)}/L^{(2)}$
and $L^{(\infty)}/L^{(3)}$ behave in the limits $\lambda \to 0$
and $\lambda \to \infty$, though $L/\omega_0$ tends to zero in the
both limits. Consider first the limit $\lambda \to \infty$. From
Eqs.~(\ref{TLfd}) and (\ref{TLft}) the following asymptotic
relations are easily derived for the fast-chaotic doublet and
triplet cases, respectively:

\begin{equation}
\frac{T^{(2)}_\mathrm{L}}{T_\mathrm{pert}} =
\frac{\mu_\mathrm{libr} + 1}{2 C_h (\mu_\mathrm{libr} + 2)} \lambda^2
\approx \frac{5}{12 C_h} \lambda^2
\approx 0.521 \lambda^2
\label{T2Tp}
\end{equation}

\noindent and

\begin{equation}
\frac{T^{(3)}_\mathrm{L}}{T_\mathrm{pert}} = \frac{1}{4 C_h} \lambda^2
\approx 0.313 \lambda^2 .
\label{T3Tp}
\end{equation}

\noindent Thus

\begin{equation}
\frac{L^{(3)}}{L^{(2)}} =
2 \frac{\mu_\mathrm{libr} + 1}{\mu_\mathrm{libr} + 2}
\approx 5/3 \approx 1.67
\label{L2L3}
\end{equation}

\noindent asymptotically. As pointed out in
Subsection~\ref{lemsm}, it is expected that $L^{(\infty)} =
L^{(3)}$ at $\lambda \to \infty$; therefore, $L^{(\infty)}/L^{(2)}
\approx 1.67$ as well.

Note that the asymptotic behavior of
${T^{(3)}_\mathrm{L}}/{T_\mathrm{pert}}$, given by
Eq.~(\ref{T3Tp}), is somewhat different from the average behavior
of ${T^{(\infty)}_\mathrm{L}}/{T_\mathrm{pert}}$ (on the interval
$0 < K (= (2 \pi /\lambda)^2) < 1.1$), expressed in
Eq.~(\ref{TLinf}). Indeed, according to Eq.~(\ref{TLinf}), at
$\lambda \gtrsim 6$ one has
$T^{(\infty)}_\mathrm{L}/T_\mathrm{pert} \approx 0.190 \lambda^2$;
i.e., the coefficient at $\lambda^2$ is 1.65 times less. The
difference is explained by the fact that the linear-looking
smoothed $L(K)$ dependence for the standard map at $0 < K < 1$
actually has the slope that weakly decreases with $K$, as also
pointed out in Subsection~\ref{lemsm}.

Thus, as followed from Eq.~(\ref{L2L3}), at $\lambda \to \infty$
one expects $L^{(\infty)}/L^{(2)} \approx 1.67$; in other words,
the relative range of the Lyapunov exponent values, if $\lambda$
is large, is rather narrow: the Lyapunov exponent in the infinitet
is only about 70\% greater than that in the doublet.

The range does not seem to be so narrow at all in the opposite
(adiabatic) limit $\lambda \to 0$. Indeed, from Eqs.~(\ref{TLsd}),
(\ref{TLst}) and (\ref{TLinf}) one finds in this limit that
$L^{(\infty)}/L^{(2)} \to \infty$ (whereas $L^{(3)}/L^{(2)} \to
1$). However note that this fact is not of much importance for
applications, because $L/\omega_0 \to 0$ at $\lambda \to 0$.

Concluding this Section, let us discuss the effect of the
perturbation strength in more detail. For the perturbation
amplitudes $\varepsilon \sim \lambda^{-1}$ and above the standard
Poincar\'e--Melnikov method for calculating the effects associated
with the separatrix splitting generally requires corrections
\cite{G97,T98}. What if the perturbation is ultimately large,
i.e., $\varepsilon = 1$? In the doublet case, the perturbation is
completely asymmetric ($\eta = 0$) and for this reason, according
to \cite{G97}, the correction is zero. For the triplet of
arbitrary asymmetry, the correction factor $R$ to the separatrix
map parameter $W$ for system~(\ref{h}), according to the Sim\'o
hypothetical formula \cite{G97}, is $\vert R(x) \vert = \left|
{\sinh (x) \over x} \right|$, where $x \equiv (2 \varepsilon_1
\varepsilon_2)^{1/2} = {(2 a b)^{1/2} \over \cal{F}}$. (The value
of $x$ may be either real or imaginary, depending on the signs of
$a$ and $b$. The value of $W$ is corrected by means of multiplying
it by $R$; i.e., the product $R W$ is used instead of $W$.) In the
symmetric triplet case, $\eta = 1$ and the correction factor is
$R(\sqrt{2}) \approx 1.3683$. Thus the correction factor in the
case of three equally-sized equally-spaced resonances is
significantly smaller than that in the case of infinitely many
equally-sized equally-spaced resonances, where $R \approx 2.2552$
(see Subsection~\ref{lemsm}).

\begin{figure}[t!]
\centering
\includegraphics[width=0.7\textwidth]{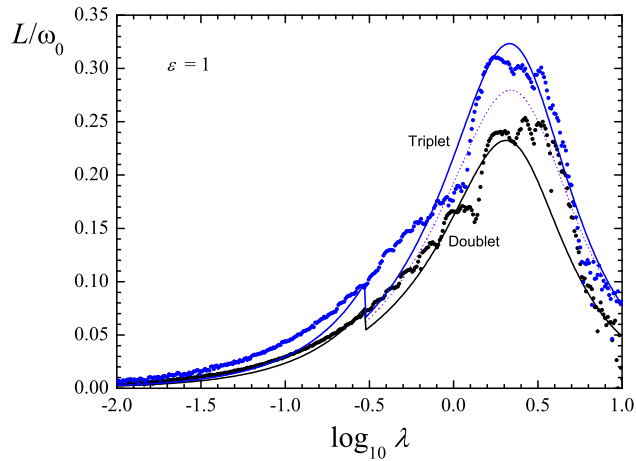} \\
\caption{The $\lambda$ dependences of the maximum Lyapunov
exponent, normalized by $\omega_0$, for the cases of equally-sized
doublet and equally-sized equally-spaced triplet. The dots show
numerical-experimental data, and the curves show theoretical
functions. The lower solid curve represents the separatrix map
theory (given by Eqs.~(\protect\ref{TLfd}) and
(\protect\ref{TLsd})) for the doublet. The upper solid curve
represents the separatrix map theory for the triplet (given by
Eqs.~(\protect\ref{TLft}) and (\protect\ref{TLst})) with the
Sim\'o correction; the middle thin dashed curve is the same but
without the Sim\'o correction. } \label{msm_fig4}
\end{figure}

Fig.~\ref{msm_fig4} shows the $\lambda$ dependences of the maximum
Lyapunov exponent, normalized by $\omega_0$, for the cases of
equally-sized doublet and equally-sized equally-spaced triplet.
The dots show numerical-expe\-rimental data, and the solid curves
show theoretical functions. The lower solid curve represents the
separatrix map theory (given by Eqs.~(\protect\ref{TLfd}) and
(\protect\ref{TLsd})) for the doublet. The upper solid curve
represents the separatrix map theory for the triplet (given by
Eqs.~(\protect\ref{TLft}) and (\protect\ref{TLst})) with the
Sim\'o correction; the middle thin dashed curve is the same but
without the Sim\'o correction. One can see that taking into
account the Sim\'o correction provides a much better fit to the
numerical data, as expected.

Note that the resonances in the infinitet (the case of the
standard map) start to overlap, on decreasing $\lambda$, at
$K_\mathrm{G} \approx 0.9716$ \cite{C79,Meiss92}, i.e., at
$\lambda = 2 \pi / \sqrt{K_\mathrm{G}} \approx 6.37$ (see
Section~\ref{sxmap}). Therefore, the ranges in $\lambda$ in
Figs.~\ref{msm_fig3a}--\ref{msm_fig4} almost completely correspond
to the overlap condition, except at $\lambda \gtrsim 6.4$, i.e.,
at $\log_{10} \lambda \gtrsim 0.8$.

The basic conclusion following from our numerical experiments,
described in this Section, is that at any given value of the
adiabaticity parameter $\lambda$ (which controls the degree of
interaction/overlap of resonances in the resonance multiplet) the
value of the maximum Lyapunov exponent in the multiplet of
equally-spaced equally-sized resonances is minimal in the doublet
case and maximal in the infinitet case. This is consistent with
the separatrix map and standard map theories: as it is clear from
Fig.~\ref{msm_fig3}, the theoretical curves for the doublet and
infinitet serve as the lower and upper bounds for all our
numerical data on the Lyapunov exponents in the multiplets.

\section{An example of application}

Resonances with planets are ubiquitous in the motion of asteroids
(see, e.g., \cite{Morbi02}); of particular interest are the
so-called mean motion resonances with Jupiter, i.e., the
resonances between orbital periods of an asteroid and Jupiter
(note that Jupiter is the largest planet in the Solar system and
is closest, among the giant planets, to the main asteroid belt;
``mean motion'' is the mean orbital frequency). The Hamiltonian of
the motion of an asteroid with negligible mass in the
gravitational field of the Sun and Jupiter, in the plane of
Jupiter's orbit, in the vicinity of a high-order mean motion
resonance with Jupiter, can be approximated \cite{HM96,MH97} in
the perturbed pendulum model as

\begin{equation}
H = {1\over2} \beta \Lambda^2 - \sum_{p=0}^{q}\phi_{k+q,\, k+p,\,
k}\cos(\psi - p \omega_1), \label{HM}
\end{equation}

\noindent where $\Lambda = \Psi - \Psi_\mathrm{res}$, $\Psi =
(\mu_1 a)^{1/2}/k$, $\Psi_\mathrm{res} = (\mu_1^2/(k^2 (k+q)
n_\mathrm{J}))^{1/3}$. The leading resonant angle $\psi \equiv k l
- (k+q) l_\mathrm{J}$, where $l$ and $l_\mathrm{J}$ are the mean
longitudes of an asteroid and Jupiter. (Definitions of the orbital
elements see, e.g., in \cite{Morbi02}.) The action-like variable
$\Lambda$ is canonically conjugated to $\psi$. The quantity
$\beta={3k^2} / a^2$ is assumed to be a constant parameter; $a$
and $e$ are asteroid's semimajor axis and eccentricity; $\omega_1
\equiv - \varpi$, i.e., $\omega_1$ is minus the longitude of
asteroid's perihelion; its time derivative is assumed to be
constant.  The units are chosen in such a way that the
gravitational constant, the total mass (Sun plus Jupiter), and
Jupiter's semimajor axis $a_\mathrm{J}$ are all equal to 1;
Jupiter's mass in the total mass units is $\mu = 1/1047.355$;
$\mu_1 = 1-\mu$. Jupiter's mean motion $n_\mathrm{J} = 1$; i.e.,
the adopted time unit is equal to ${1 \over 2 \pi}$th part of
Jupiter's orbital period.

The integer non-negative numbers $k$ and $q$ define the resonance:
the ratio $(k+q)/k$ is equal to the ratio of mean motions of an
asteroid and Jupiter in the exact resonance; $q$ is the resonance
order. According to Eq.~(\ref{HM}), the mean motion resonance
$(k+q)/k$ splits in a cluster of $q+1$ subresonances $p = 0, 1,
\ldots, q$. For the coefficients of the resonant terms one has

\vspace{-3mm}

\begin{equation}
|\phi_{k+q,\, k+p,\, k}| \approx {\mu \over {q \pi a_\mathrm{J}}}
{q \choose p} \left(\epsilon \over 2 \right)^p
\left(\epsilon_\mathrm{J} \over 2 \right)^{q-p}, \label{coeff}
\end{equation}

\noindent where $\epsilon \equiv {{e a_\mathrm{J}} /
{(a_\mathrm{J} - a)}}$, $\epsilon_\mathrm{J} \equiv {{e_\mathrm{J}
a_\mathrm{J}} / {(a_\mathrm{J} - a)}}$. Jupiter's current
eccentricity is $e_\mathrm{J}=0.048$. The frequency of
small-amplitude oscillations on subresonance $p$ is

\vspace{-3mm}

\begin{equation}
\omega_0 = (\beta|\phi_{k+q,\, k+p,\, k}|)^{1/2} \approx
{a_\mathrm{J}\over{a_\mathrm{J} - a}} n_\mathrm{J} \left( \mu_1
\mu {{4q} \over {3\pi}} {q \choose p} \left(a \over
{a_\mathrm{J}}\right) \left(\epsilon \over 2 \right)^p
\left(\epsilon_\mathrm{J} \over 2 \right)^{q-p} \right)^{1/2},
\label{om0}
\end{equation}

\noindent and the perturbation frequency is

\begin{equation}
\Omega = \dot \omega_1 \approx \frac{\mu_1 \mu}{2 \pi}
n_\mathrm{J} \left(\frac{a}{a_\mathrm{J}}\right)^{1/2}
\left(\frac{a_\mathrm{J}}{a_\mathrm{J} - a}\right)^2, \label{Om}
\end{equation}

\noindent cf.~\cite{HM96, MH97}.

As an example we take asteroid 522~Helga, which is famous to
exhibit ``stable chaos'' \cite{MN92,MN93,TVH00,S07IAU}: i.e., its
computed Lyapunov time is rather small ($\sim 7000$~yr), but
numerical experiments do not reveal any gross changes of its orbit
on cosmogonic time scales. Helga is known to be in the 12/7 mean
motion resonance with Jupiter. We take necessary data on $a$, $e$,
and the perihelion frequency $g = \dot \varpi$ for this asteroid
in the ``numb.syn'' catalogue \cite{KM03,KM12} of the AstDyS web
service\footnote{http://hamilton.dm.unipi.it/astdys/}. The value
of $T_\mathrm{pert}$ is defined by the value of $g$; thus one
finds $T_\mathrm{pert} = 6700$~yr.

To apply the separatrix map theory, one should identify the
guiding subresonance in the multiplet. As such, it is natural to
choose the subresonance that has the maximum amplitude (i.e., the
maximum value of $|\phi_{k+q,\, k+p,\, k}|$). We find that the
guiding subresonance in the sextet is the third one ($p = 2$),
consequently the perturbing neighbors have numbers $p = 1$ and 3.
Thus we find the separatrix map parameters: $\lambda=
\Omega/\omega_0 = 2.32$, $\eta = 0.81$. Therefore, we model the
multiplet by a fast-chaotic triplet. The relative strength of
perturbation is rather strong: $\varepsilon = 0.79$. Applying
Eq.~(\ref{TLft}), one has $T_L \approx 9800$~yr.

On the other hand, the standard map theory gives an estimate for
the Lyapunov time from below. According to Eq.~(\ref{Kla}), $K =
(2 \pi /\lambda)^2$; thus one has for 522~Helga: $K \approx 7.3$,
and, as follows from Eq.~(\ref{TLinf}), $T_L \approx 5100$~yr.

Values of the Lyapunov time, computed in integrations in the full
(accounting for perturbations from all major planets) problem are
6900~yr \cite{MN93} and 6860~yr (AstDyS). Obviously, the standard
map theory is closer to these ``actual'' values. This is because
the number of resonances in the multiplet is large and the
relative strength of perturbation $\varepsilon$ is not far from 1,
i.e., to the value characteristic for the standard map
Hamiltonian.

\section{Conclusions}
\label{concl}

In this article, the problem of estimating the maximum Lyapunov
exponents of the motion in a multiplet of interacting resonances
has been considered for the case when the resonances have
comparable strength. The corresponding theoretical approaches have
been considered for the multiplets of two, three, and infinitely
many interacting resonances  (i.e., doublets, triplets, and
``infinitets''). The analysis has been based on the theory of
separatrix and standard maps. We have introduced a ``multiplet
separatrix map'', valid for description of the motion in the
resonance multiplet under certain conditions.

The separatrix map approach is suitable for the multiplet of any
number of resonances, when their interaction is weak or moderate
(i.e., the separation of resonances with respect to their sizes is
large enough), as well as for the multiplet of two or three
resonances (doublet or triplet), when the degree of interaction is
arbitrary, including the case of strong overlap. The standard map
approach is suitable for the multiplet of a large number of
equally-sized equally-spaced resonances with arbitrary degree of
inter\-action/over\-lap.

We have presented explicit analytical formulas for the Lyapunov
times for the following five generic resonance multiplet types:
fast-chaotic resonance triplet, fast-chaotic resonance doublet,
slow-chaotic resonance triplet, slow-chaotic resonance doublet,
and, for both cases of fast and slow chaos, infinitet of
equally-sized equally-spaced resonances. Good performance of the
presented analytical formulas in the domains of their validity has
been demonstrated by means of comparison with direct numerical
integrations of the original Hamiltonian systems.

In numerical experiments we have shown that, at any given value of
the adiabaticity parameter $\lambda$, the value of the maximum
Lyapunov exponent in the multiplet of equally-spaced equally-sized
resonances is minimal in the doublet case and maximal in the
infinitet case. This is consistent with the developed theory.

An example of application of the developed theory has been given,
concerning asteroidal dynamics in high-order mean motion
resonances with Jupiter.

\section*{Acknowledgements}

The author is thankful to the referee for useful remarks. This
work was supported in part by the Programmes of Fundamental
Research of the Russian Academy of Sciences ``Fundamental Problems
in Nonlinear Dynamics'' and ``Fundamental Problems of the Solar
System Studies and Exploration''. The computations were partially
carried out at the St.~Petersburg Branch of the Joint
Supercomputer Centre of the Russian Academy of Sciences.

\end{document}